\documentclass[%
 reprint,
showpacs,preprintnumbers,
 amsmath,amssymb,
 aps,
prb,
]{revtex4-1}
\usepackage{amsmath}
\usepackage{stmaryrd}
\usepackage{txfonts}
\usepackage{amssymb}
\usepackage{mathrsfs}
\usepackage{epsfig}
\usepackage[colorlinks=true, linkcolor=blue, citecolor=blue, urlcolor=blue]{hyperref} 
\usepackage{graphicx}% Include figure files
\usepackage{dcolumn}% Align table columns on decimal point
\usepackage{bm}% bold math
\usepackage{color}
\usepackage{ulem}

\begin{document}

\title{Bubble and skyrmion crystals in frustrated magnets with easy-axis anisotropy}% Force line breaks with \\

\author{Satoru Hayami,$^{1,2}$ Shi-Zeng Lin,$^1$ and Cristian D. Batista$^{1,3}$}
\affiliation{
$^1$Theoretical Division, T-4 and CNLS, Los Alamos National Laboratory, Los Alamos, New Mexico 87545, USA \\
$^2$Department of Physics, Hokkaido University, Sapporo 060-0810, Japan\\
$^3$Quantum Condensed Matter Division and Shull-Wollan Center, Oak Ridge National Laboratory, Oak Ridge, TN 37831, USA
 }

 
\begin{abstract}
We clarify the conditions for the emergence of multiple-${\bf Q}$ 
structures out of  lattice  and easy-axis spin anisotropy in frustrated magnets. By considering 
magnets whose exchange interaction has multiple global minima in momentum space, we  find that both types of anisotropy stabilize triple-${\bf Q}$ orderings. Moderate anisotropy leads to a magnetic field-induced skyrmion crystal, which evolves into a bubble crystal for increasing spatial and spin anisotropy. The bubble crystal exhibits a quasi-continuous (devil's staircase) temperature dependent ordering wave-vector, characteristic of the competition between frustrated exchange and strong
easy-axis anisotropy.
\end{abstract}
\maketitle

\section{Introduction}
\label{sec:Introduction}
Helical spin states were originally observed in rare-earth and other itinerant magnets,~\cite{yoshimori1959new,Kaplan_PhysRev.124.329,Elliott_PhysRev.124.346}   whose localized magnetic moments interact via 
the Ruderman-Kittel-Kasuya-Yosida (RKKY) interaction, $-J^2 \sum_{\bf q} \chi^0_{\bf q} {\bf S}_{\bf q} \cdot {\bf S}_{\bf -q}$.~\cite{Ruderman,Kasuya,Yosida1957} 
Because this interaction is mediated by conduction electrons coupled to the local moments by an exchange $J$, it selects ordering wave-vectors, ${\bf Q}$, 
which maximize the electronic magnetic susceptibility  $\chi^0_{\bf q}$. 
However, as it was recently recognized,~\cite{Martin_PhysRevLett.101.156402,Akagi_JPSJ.79.083711,Kato_PhysRevLett.105.266405,Akagi_PhysRevLett.108.096401,Barros_PhysRevB.88.235101,Hayami_PhysRevB.90.060402,Ozawa15} single-${\bf Q}$ helical orderings  can  become unstable towards multiple-${\bf Q}$ modulated structures whenever 
$\chi^0({\bf q})$ has global maxima at different symmetry related wave-vectors ${\bf Q}_{\nu}$. This instability has its
roots in the relatively large strength of four and higher-spin interactions, which arise from tracing out conduction electrons {\it beyond the RKKY level.}~\cite{Hayami_PhysRevB.90.060402,Ozawa15}

From a real-space viewpoint, low-symmetry wave-vectors of helical 
orderings arise from competition between exchange interactions. This competition does not require  
long-range (power-law decay) interactions, like the RKKY coupling of inter-metallic systems. Mott insulators can 
also exhibit competing (short-range) exchange interactions that favor helical ordering.~\cite{nakatsuji2005spin,regnault1982inelastic,day1981neutron} However, unlike the case of  itinerant magnets, 
four and higher-spin interactions are usually weak in these systems. It is then natural to look
for alternative ways of stabilizing multiple-${\bf Q}$ structures in high-symmetry 
frustrated Mott insulators, whose exchange interaction in momentum space, $J({\bf q})$, has multiple global minima.

The triangular lattice (TL) provides a simple realization of a high-symmetry system with six equivalent orientations for the helix. This symmetry allows for an anharmonic interaction between triple-${\bf Q}$ modulations 
and the uniform magnetization induced by an external field because ${\bf Q}_1+{\bf Q}_2+{\bf Q}_3=0$.~\cite{Garel_PhysRevB.26.325} 
Indeed, Monte Carlo (MC) simulations of a frustrated $J_1$-$J_3$ classical Heisenberg model on a TL revealed a skyrmion crystal at finite temperature and magnetic field values.~\cite{Okubo_PhysRevLett.108.017206}  
The origin of this phase is quite different from the skyrmion crystals (SC's) that emerge in chiral magnets out of the  competition between Dzyaloshinskii-Moriya and ferromagnetic exchange interactions.~\cite{Bogdanov89,Bogdanov94,rossler2006spontaneous}
Moreover, because the chiral and U(1) symmetries are spontaneously broken in non-chiral magnets, 
their metastable single skyrmions have different properties.~\cite{leonov2015multiply,lin2015ginzburg}  

Single-${\bf Q}$ orderings are favored by the exchange interactions at $T=0$ because multi-${\bf Q}$ orderings
are accompanied by  higher harmonics required to preserve the spin normalization $\mathbf{S}_i^2=1$.
Quantum or thermal fluctuations make the longitudinal spin stiffness finite and can heavily suppress it
near quantum critical points or thermodynamic phase transitions. Indeed, triple-${\bf Q}$ magnetic orderings, such as vortex and SC's, have been reported for both regimes.~\cite{Kamiya_PhysRevX.4.011023,Wang_PhysRevLett.115.107201,Okubo_PhysRevLett.108.017206,Seabra15}
Fluctuations then play an important role in the subtle competition between 
single-${\bf Q}$ and different multi-${\bf Q}$ orderings. Easy-axis anisotropy is also expected to favor 
multi-${\bf Q}$ orderings, as it was recently shown by means of purely classical $T=0$ variational calculations.~\cite{leonov2015multiply} In this letter we use unbiased MC simulations of the $J_1$-$J_2$ and 
$J_1$-$J_3$  triangular Heisenberg models with easy-axis anisotropy to demonstrate that thermal fluctuations
 modify substantially the $T=0$ phase diagram.  

\begin{figure}[t]
\begin{center}
\includegraphics[width=1.0 \hsize]{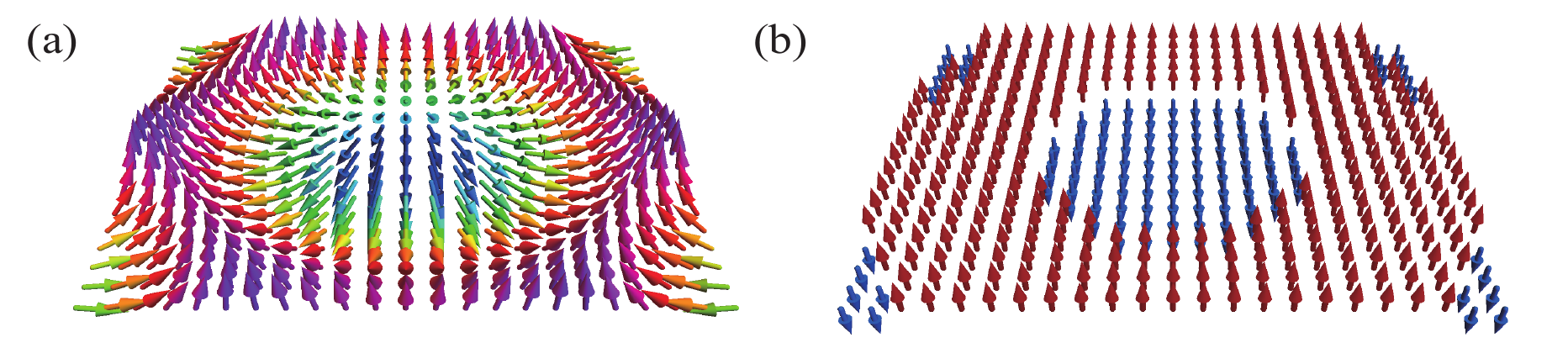} 
\caption{
\label{Fig:ponti}
(Color online)
Schematic views of (a) a noncoplanar skyrmion texture and (b) a collinear bubble. Triangular crystals of 
these structures are induced by magnetic field and easy-axis anisotropy in high-symmetry frustrated magnets.
}
\end{center}
\end{figure}

By combining  MC simulations with variational $T=0$ calculations, we clarify the range of stability of the skyrmion crystal found in Ref.~\onlinecite{Okubo_PhysRevLett.108.017206}. In absence of an easy-axis anisotropy, the six-fold spatial anisotropy
plays a crucial role in the stabilization of the skyrmion crystal [Fig.~\ref{Fig:ponti}(a)]. Indeed, the skyrmion crystal phase disappears for a small
$Q$ and it only reappears for  moderate easy-axis anisotropy. The  field-induced skyrmion crystal evolves  into a bubble crystal (BC) [see Fig.~\ref{Fig:ponti}(b)] for larger spatial and spin anisotropies. This triple-${\bf Q}$ collinear state exhibits a devil's staircase-like temperature-dependent ordering wave-vector, characteristic of the competition between frustrated exchange and easy-axis anisotropy.

The rest of the paper is organized as follows. 
After introducing a frustrated Heisenberg model on a TL in Sec.~\ref{sec:Model}, we show in Sec.~\ref{Isotropic Hamiltonian}
that in absence of spin anisotropy this model exhibits a skyrmion crystal phase only above a critical value of $Q$. 
In Sec.~\ref{sec:Multi-Q Orderings by a Single-Ion Anisotropy} we demonstrate 
that a single-ion easy-axis anisotropy naturally leads to multi-$\mathbf{Q}$ magnetic orderings irrespective of the magnitude of the ordering vector. 
Sec.~\ref{sec:Numerical Analysis} includes  a $T=0$ variational analysis and finite-$T$ MC simulations for relatively small wave-vectors $Q$.
These results are combined to produce different phase diagrams as a function of temperature, magnetic field and single-ion anisotropy. 
In particular, we show that the easy-axis anisotropy gives rise to multiple-$Q$ states, such as skyrmion  and bubble crystal phases.
In Sec.~\ref{sec:Large-Q} we provide a similar analysis for large ${\bf Q}$-values.
A summary of the results is presented in Sec.~\ref{sec:Summary}. 

\section{Model}
\label{sec:Model}

We consider a frustrated Heisenberg model on a TL: 
\begin{eqnarray}
\label{eq:Ham}
\mathcal{H}= \sum_{\langle i,j \rangle} J_{ij} \mathbf{S}_i \cdot \mathbf{S}_j - H \sum_{i}  S^z_i - A \sum_i (S^z_i)^2.
\end{eqnarray}
The classical moments, $\mathbf{S}_i$, have a fixed magnitude $\mathbf{S}_i^2=1$. 
The first term is the isotropic exchange interaction, including
nearest-, second nearest-, and third nearest-neighbor couplings, $J_1$, $J_2$, and $J_3$, respectively. 
The ferromagnetic interaction, $J_1=-1$, will be our unit of energy and the lattice constant, $a$, will be adopted as the
unit of length. Frustration arises from the antiferromagnetic 
nature of the further neighbor interactions $J_2>0$ and $J_3>0$. 
The second and third terms represent the Zeeman coupling to an external magnetic field and the easy-axis ($A>0$) spin anisotropy, respectively. $\mathcal{H}$ is invariant under the space-group of the TL and 
under the U(1) group of  global spin rotations along the field-axis.

Below the saturation field, $H_{\rm sat}$,  the ground state of $\mathcal{H}(A=0)$ is the conical spiral: $\mathbf{S}_i =(\sin\theta \cos(\mathbf{Q}\cdot \mathbf{r}_i),\ \sin\theta\sin(\mathbf{Q}\cdot \mathbf{r}_i),\ \cos\theta)$ with $Q=|\mathbf{Q}| = (2/\sqrt{3}) \cos^{-1} \left[ (1+J_1/J_2)/2 \right]$ for the $J_1$-$J_2$ model ($J_3=0$) and $Q=2 \cos^{-1} \left[ (1+\sqrt{1-2J_1/J_3})/4 \right]$ for the $J_1$-$J_3$ model ($J_2=0$).  
In each case there are six possible ordering wave-vectors, $\pm{\bf Q}_{\nu}$ $({\nu=1,2,3})$, because of the C$_6$ symmetry of the TL. These  vectors are parallel to the nearest (next-nearest) neighbor bond directions for the  $J_1$-$J_3$ ($J_1$-$J_2$) model. 
The canting angle $\theta$  is given by $\cos\theta=H/H_{\rm sat}$ with $H_{\rm sat}=J (\mathbf{Q}) - J (\mathbf{0}), $ and 
$J (\mathbf{q})=\sum_{\boldsymbol \delta} J_{\boldsymbol \delta} e^{i {\bf q} \cdot \boldsymbol \delta }$ (${\boldsymbol \delta}$ is the relative
vector between neighboring sites). From this relationship, we obtain  that $H_{\rm sat}\propto Q^4$ for $Q\ll 1$, i.e., near the Lifshitz transition
to the commensurate $Q=0$ ferromagnetic state.

\section{Isotropic Spin Interactions}
\label{Isotropic Hamiltonian}

We start by considering isotropic spin interactions ($A=0$) in order to isolate the effect of the six-fold lattice anisotropy.
This anisotropy appears upon expanding $J (\mathbf{q})$ up to sixth order in $q_x$ and $q_y$. For the $J_1$-$J_3$ model we have
\begin{eqnarray}
\label{eq:expandJ}
J (\mathbf{q}) = &-&6 (J_1+J_3) + \frac{3}{2} (J_1 + 4 J_3) q^2
-\frac{3}{32}(J_1+16 J_3) q^4 \nonumber \\ 
&+&\frac{1}{384}(J_1+64J_3) q^6 + \frac{1}{3840} (J_1+64 J_3) q^6 \cos 6 \phi , 
\nonumber \\
\end{eqnarray}
where $\mathbf{q}=(q \cos \phi, q \sin \phi)$.  

The  thermodynamic phase diagram of the $J_1$-$J_2$ and the $J_1$-$J_3$ is obtained 
from unbiased MC simulations based on the Metropolis algorithm and the over-relaxation method.  The lattices used for these simulations have $N=L^2$ spins and  periodic boundary conditions. 
The target temperature is reached by simulated annealing over $10^5$-$10^6$ MC sweeps (MCS) and  $10^5$-$10^7$ MCS measurements are performed after equilibration. Statistical errors are estimated by taking averages over $3$-$16$ independent runs.

According to our MC simulation of ${\cal H}$ on $L=75$, $98$, $100$ $120$ lattices, the conical spiral  is
the only ordered phase for small enough $Q$. For the  $J_1$-$J_3$ model, the skyrmion crystal phase only appears above $Q^{\rm c}=1.980(4)$, which corresponds to $J_2^{\rm c}/|J_1|=1.4027(138)$. For the  $J_1$-$J_3$ model, we obtain
$Q^{\rm c}=1.648(2)$, which corresponds to $J_3^{\rm c}/|J_1|=1.0256(53)$. These results indicate  that the locking potential, 
which grows as $Q^6$ and forces three helices to propagate along the  principal axes of the TL, has to reach a critical value 
to stabilize the skyrmion crystal phase in isotropic magnets. As we will see in the next sections, this condition is no longer required in the presence of 
a moderate easy-axis anisotropy.

\section{Multi-$\mathbf{Q}$ Instability Induced by a Single-Ion Anisotropy}
\label{sec:Multi-Q Orderings by a Single-Ion Anisotropy}

The purpose of this Section is to demonstrate that a finite easy-axis anisotropy is enough to stabilize multi-${\bf Q}$ orderings.
To this end we will perform a stability analysis of the single-${\bf Q}$ conical spiral phase based on the following deformation:~\footnote{Ref.~22 provides a similar argument but
only for the continuum limit ($Q \ll 1$).} 
\begin{eqnarray}
S^x_j &=& \sqrt{\sin^2{\!\tilde \theta}- \Delta^2_2 } \cos{({\bf Q}_{1} \cdot {\bf r}_j )} + \Delta_2 \cos{({\bf Q}_{2} \cdot {\bf r}_j )},
\nonumber \\
S^y_j &=& \sqrt{\sin^2{\! \tilde \theta}- \Delta^2_2 } \sin{({\bf Q}_{1} \cdot {\bf r}_j )} - \Delta_2 \sin{({\bf Q}_{2} \cdot {\bf r}_j )},
\nonumber \\
S^z_j &=&   \sqrt{\cos^2{\! \tilde \theta} - 2 \Delta_2 \sqrt{\sin^2{\!\tilde \theta} - \Delta^2_2} \cos{{\bf Q}_3} \cdot {\bf r}_j},
\label{con2}
\end{eqnarray}
where the amplitude of the ${\bf Q}_2$ component,  $\Delta_2$, is a variational parameter  
and $\cos{\tilde{\theta}}$ is determined below [see Eq.~(\ref{costheta})]. 

We will demonstrate that the energy of the variational state given in Eq.~(\ref{con2}) is a decreasing function of $\Delta_2$ for $\Delta_2 \ll 1$. This means that the single-${\bf Q}_1$
conical state ($\Delta_2$=0) is unstable towards the development of a second ${\bf Q}_2$ component, as long as 
the magnetic field, $H$, and the easy-axis anisotropy, $A$, are non-zero. 
We will then expand the 
total energy per site, $E(\Delta_2)$, to fourth order in $\Delta_2$. In general, the total energy per site of an arbitrary state is given by
\begin{eqnarray}
E = \langle \mathcal{H} \rangle = \frac{1}{N} \sum_{\bf q} J ({\bf q}) |\langle {\bf S}_{\bf q}  \rangle |^2
- \frac{H \langle S^z_{\bf 0} \rangle}{\sqrt{N}}
- \frac{A}{N} \sum_j \langle (S^z_j)^2 \rangle, 
\label{energy}
\end{eqnarray}
with
\begin{eqnarray}
{\bf S}_{\bf q} = \frac{1}{\sqrt{N}} \sum_{j} {\bf S}_j e^{i {\bf q} \cdot {\bf r}_j}.
\end{eqnarray}
For the state under consideration, we have:
\begin{widetext}
\begin{eqnarray}
S^z_j = \cos{\tilde \theta} \left [ 1 - x \cos{{\bf Q}_3 \cdot {\bf r}_j} - \frac{x^2}{2} \cos^2{\! {\bf Q}_3 \cdot {\bf r}_j}
- \frac{x^3}{2} \cos^3{\! {\bf Q}_3 \cdot {\bf r}_j} - \frac{5x^4}{8} \cos^4{\! {\bf Q}_3 \cdot {\bf r}_j} + {\cal O}(\Delta^5_2) \right],
\end{eqnarray}
\end{widetext}
with
\begin{equation}
x= \frac{\Delta_2}{\cos^2{\! \tilde \theta}} \sqrt{\sin^2{\! \tilde \theta} - \Delta^2_2}.
\end{equation}
We will choose ${\tilde \theta}$, such that
\begin{equation}
\cos{\theta} \! = \! \cos{\tilde \theta} \; [1- \Delta^2_2 (\sin^2{\!\tilde \theta} - \Delta^2_2) / (4 \cos^4{\! \tilde \theta} )].
\label{costheta}
\end{equation}
For this choice of ${\tilde \theta}$ we have:
\begin{eqnarray}
 \frac{|\langle S^z_{\bf 0}\rangle|}{\sqrt{N}}  &=& \cos{\theta} - \frac{15}{64} \cos{\tilde \theta} x^4 + {\cal O}(\Delta^5_2)
\nonumber \\
\frac{|\langle S^z_{{\bf Q}_3}\rangle|^2}{N}  &=& \cos^2{\! \tilde \theta} \left [ \frac{ x^2}{2}
+ \frac{3x^4}{8} \right]  +{\cal O}(\Delta^6_2) 
\nonumber \\
\frac{|\langle S^z_{2{\bf Q}_3}\rangle|^2}{N}  &=&  \cos^2{\! \tilde \theta} \frac{x^4}{32} + {\cal O}(\Delta^6_2)
\nonumber \\
 \sum_{\nu=1,2; \mu=x,y}  \frac{|\langle S^\mu_{{\bf Q}_{\nu}}\rangle|^2}{N} &=&  \sin^2{\! \tilde{\theta}}. 
\end{eqnarray}
By adding the different contributions to Eq.~\eqref{energy}, we obtain:
\begin{eqnarray}
E(\Delta_2) - E(\Delta_2=0) &= &\left [  \frac{J(2 {\bf Q}_3) - J ({\bf 0})}{32} + \frac{9 A}{8}\right ] x^4 \cos^2{\! \tilde \theta} \nonumber \\
&-& \frac{A x^2}{2} \cos^2{\! \tilde \theta}  + {\cal O}(\Delta^5_2),
\label{energy2}
\end{eqnarray}
where we have used that $H = 2 \cos{\tilde \theta} [J({\bf 0}) - J({\bf Q}_{\nu})]$ to zeroth order in $\Delta_2$.
It is clear from this expression that the energy is a decreasing function of $\Delta_2$ for small enough $\Delta_2$.
In particular, if we assume that $A \ll J(2 {\bf Q}_3) - J ({\bf 0}) $, we can minimize \eqref{energy2} as a function of $x$ to obtain.
\begin{equation}
x^2 = \frac{8A}{J(2 {\bf Q}_3) - J ({\bf 0)}},
\end{equation}
implying that 
\begin{equation}
\label{Delta2}
\Delta_2 \simeq \frac{\cos^2{\theta} \sqrt{8A}}{\sin{\theta} \sqrt{J(2 {\bf Q}_3) - J ({\bf 0)}}}.
\end{equation}

Thus, we find that the single-${\bf Q}$ conical state is unstable toward the multi-${\bf Q}$ deformation. 
A key observation is that the modulation of the $z$ spin component, required to preserve the constraint ${\bf S}^2_i =1$,
has a very small exchange energy cost in a C$_6$ invariant system:
to linear order in $\Delta_2$,  the $z$-component is modulated by the third wave-vector ${\bf Q}_3$, which still minimizes $J (\mathbf{q})$. This is so because  the C$_6$ symmetry of the TL guarantees that ${\bf Q}_1 + {\bf Q}_2 + {\bf Q}_3=0$. Therefore, the exchange {\it energy cost} of the higher harmonics produced by the normalization condition is proportional to $\Delta_2^4 [ J(2 {\bf Q}_3)- J({\bf 0})]$, while the anisotropy  {\it energy gain} 
produced by the same modulation  is proportional to $- A \Delta^2_2$, as it is shown in Eq.~(\ref{energy2}). In the end, this leads to 
 $\Delta_2 \propto \sqrt{A/[ J(2 {\bf Q}_3)- J({\bf 0})]}$ for $\Delta_2 \ll 1$ or $A \ll |J(2 {\bf Q}_{\nu})|$, as obtained in Eq.~(\ref{Delta2}).  
 Finally, it is interesting to note that a double-${\bf Q}$ conical state, like the one described by Eq.~(\ref{con2}), has been obtained 
below the saturation field of a {\it spatially anisotropic} TL model~\cite{Starykh14}.

\section{Small ${\bf Q}$}
\label{sec:Numerical Analysis}

In Sec.~\ref{Isotropic Hamiltonian} we showed that a critical $Q$-value is required to stabilize a skyrmion crystal for isotropic spin interactions. 
In Sec.~\ref{sec:Multi-Q Orderings by a Single-Ion Anisotropy}, we demonstrated that a single-${\bf Q}$  conical spiral phase 
is unstable towards multi-${\bf Q}$ orderings in the presence of finite easy-axis anisotropy. It is then natural to ask what is the thermodynamic
phase diagram for small $Q$-values ($Q < Q^{\rm c}$) as a function of magnetic field and  easy-axis anisotropy $A$. This is the main 
purpose of this Section. We start with a simple $T=0$ variational analysis, which is complemented by finite-$T$ MC simulations.

\begin{figure}[t!]
\begin{center}
\includegraphics[width=\columnwidth]{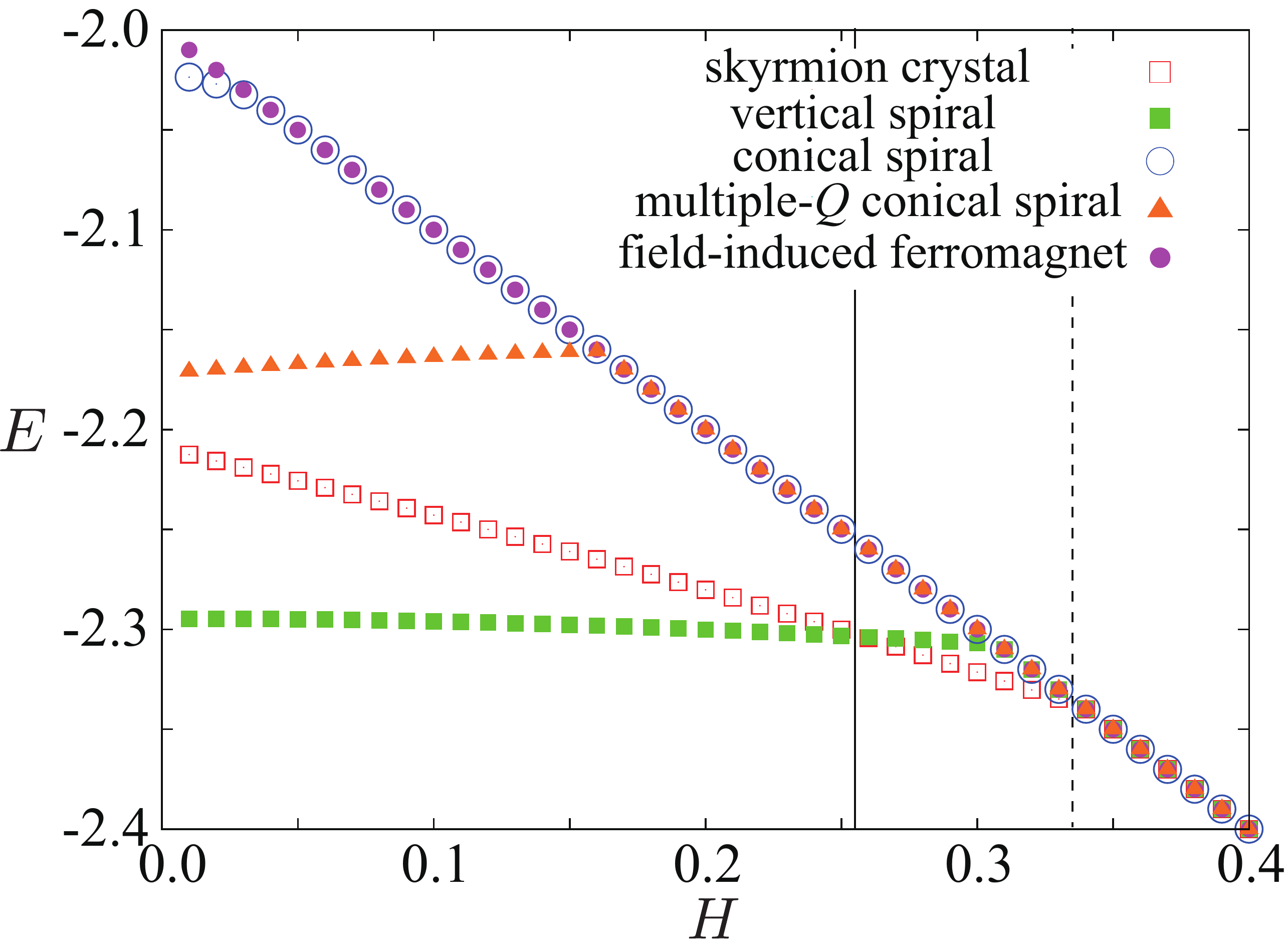} 
\caption{
\label{Fig:fig_supple_variational}
(Color online)
$H$ dependence of the energy per site of different variational states for $Q=2\pi/5$ and $A=0.5$. 
The vertical solid line marks the  phase boundary between the single-${\bf Q}$ vertical spiral and skyrmion crystal phases. The vertical dashed line
marks the phase boundary between the skyrmion crystal and fully polarized state. Here we only considered the multiple-${\bf Q}$ conical spiral with $I_x=I_y$. 
}
\end{center}
\end{figure}

\subsection{Variational Analysis }
\label{sec:Small-Q}

Here we present  a simple $T=0$ variational analysis of the $J_1$-$J_3$ model based on the following variational states:

 (1) {\it Skyrmion crystal phase:} the spin configuration is given by $\mathbf{S}=\mathbf{M}/|M|$, with
\begin{eqnarray}
{\bf M}^{x,y}_i &=& {I_{xy}} \sum_{\nu=1-3} \sin{({\bf Q}_{\nu} \cdot {\bf r}_i+ \theta_{\nu})} \;\; {\bf e}_{\nu},
\nonumber \\
M^z_i &=& m_z - {I_z} \sum_{\nu=1-3} \cos{({\bf Q}_{\nu} \cdot {\bf r}_i+ \theta_{\nu})} \;\; {\bf e}_{\nu}.
\label{sky}
\end{eqnarray}
The three unit vectors are ${\bf e}_1={\hat {\bf x}}$, ${\bf e}_2=-{\hat {\bf x}}/2 + \sqrt{3}{\hat {\bf x}}/2$, and ${\bf e}_3=-{\hat {\bf x}}/2 - \sqrt{3}{\hat {\bf x}}/2$,  $\mathbf{Q}_\nu=Q\mathbf{e}_\nu$ and
$m_z$ is the uniform spin magnetization. Higher  harmonics  are generated by the normalization condition ${\bf S}^2_i=1$. 
Without loss of generality, we take $\theta_\nu =0$ because $\mathcal{H}$ is invariant under global spin rotations along the magnetic field direction.  The variational parameters of the skyrmion crystal state \eqref{sky} are $m_z$, $I_z$, $Q_v$ and $I_{xy}$. 

(2) {\it Fully polarized state:}
\begin{eqnarray}
{\bf S}^{xy}_i = 0, \;\;\; S^z_i = 1. 
\label{ferro}
\end{eqnarray}

(3) {\it Single-$\bm{Q}$ conical spiral:}
\begin{eqnarray}
{\bf S}^{xy}_i &=& \sqrt{1-m_z^2} \;\; [\cos{({\bf Q}_{\nu} \cdot {\bf r}_i )} \;\; {\hat {\bf x}} + \sin{({\bf Q}_{\nu} \cdot {\bf r}_i )} \;\; {\hat {\bf y}}],
\nonumber \\
S^z_i &=& m_z. 
\label{con}
\end{eqnarray}

(4) {\it Single-$\bm{Q}$ vertical spiral:}  the spin configuration is given by $\mathbf{S}=\mathbf{M}/|M|$, with
\begin{eqnarray}
{\bf M}^{xy}_i &=& {I_{xy}} \cos{({\bf Q}_{\nu} \cdot {\bf r}_i )} \;\; {\hat {\bf x}}, 
\nonumber \\
M^z_i &=& m_z + {I_{z}} \sin{({\bf Q}_{\nu} \cdot {\bf r}_i )}, 
\label{hel}
\end{eqnarray}
where we have assumed that the spins rotate in the $x$-$z$ plane. Once again, the orientation of this polarization  plane is arbitrary (provided it is parallel to the $z$-axis) because $\mathcal{H}$ is U(1) invariant under global spin rotations along the $z$-axis. 

(5) {\it Multiple-$\bm{Q}$ conical spiral:} this state corresponds to Eq.~(\ref{con2}) and it is given by
$\mathbf{S}=\mathbf{M}/|M|$ with
 \begin{eqnarray}
M^{x}_i &=& {I_x} [\cos{({\bf Q}_{1} \cdot {\bf r}_i )} + \cos{({\bf Q}_{2} \cdot {\bf r}_i )}],
\nonumber \\
M^{y}_i &=&{I_y} [\sin{({\bf Q}_{1} \cdot {\bf r}_i )} - \sin{({\bf Q}_{2} \cdot {\bf r}_i )}],
\nonumber \\
M^z_i &=& m_z+{I_z} \cos({\bf Q}_3\cdot {\bf r}_i). 
\label{2q}
\end{eqnarray}
As it can be inferred from the analysis of Eqs.~(\ref{con2}) and (\ref{Delta2}), this state  has lower energy than that of the single-${\bf Q}$ conical state for finite $A$ and $H$.~\cite{leonov2015multiply}

\begin{figure}[tb!]
\begin{center}
\includegraphics[width=1.0 \hsize]{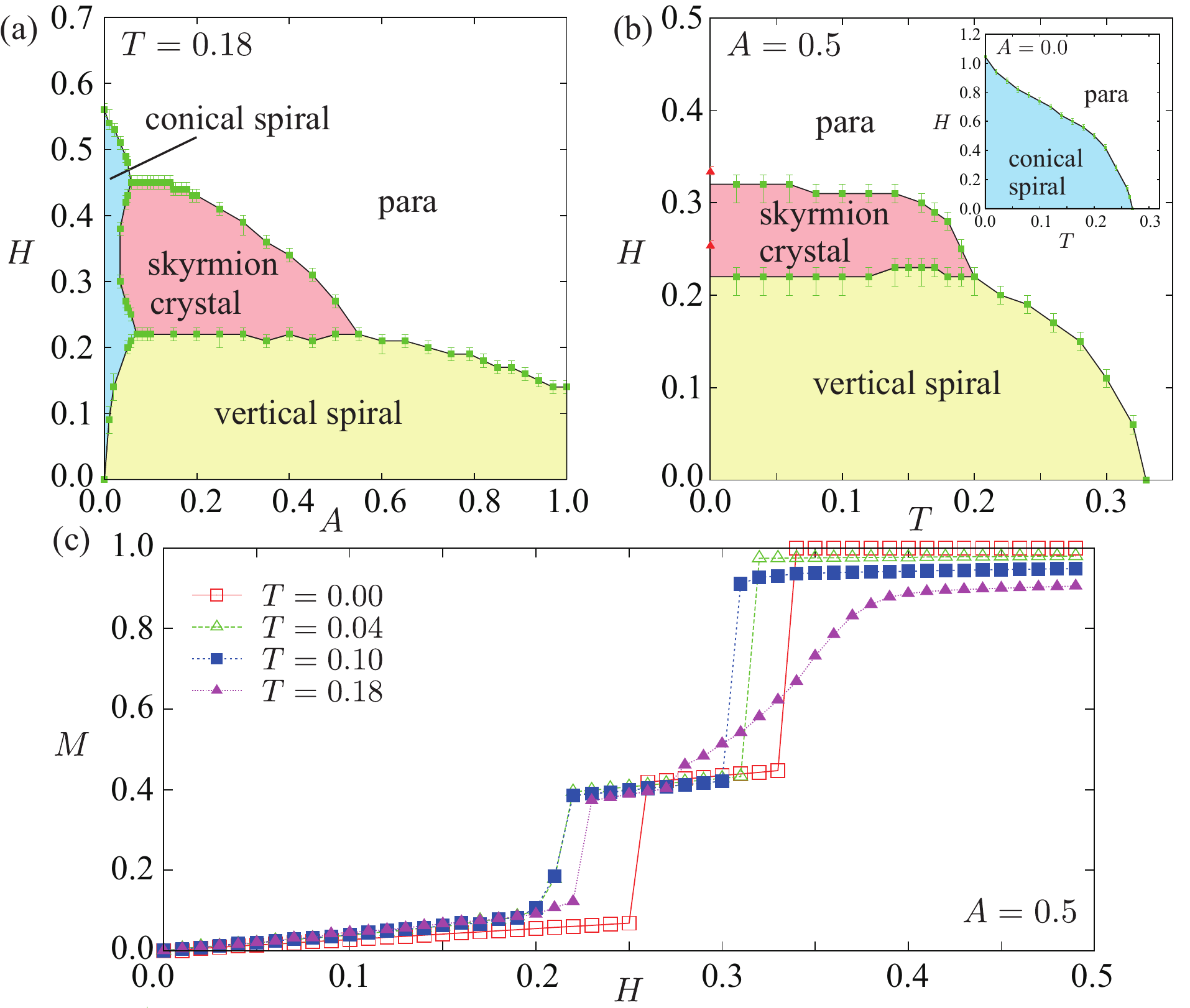} 
\caption{
\label{Fig:phase_smallq}
(Color online)
(a) $A$-$H$ phase diagram obtained from MC simulations of the $J_1$-$J_3$ model for $J_3/|J_1|=0.5$ at $T=0.18$. 
(b) $T$-$H$  phase diagram for $A=0.5$ . 
The red triangles  are determined from $T=0.0$ variational calculations. 
The inset shows the phase diagram for $A=0.0$. 
(c) Field dependence of the magnetization at different temperatures for $A=0.5$. 
The error bars are smaller than the symbol size. 
}
\end{center}
\end{figure}

Figure~\ref{Fig:fig_supple_variational} shows the $H$-dependence of the energy density of each variational state for $Q=2\pi/5$ ($J_3/|J_1|=0.5$) and $A=0.5$. 
The vertical spiral, skyrmion crystal, and fully-polarized states become stable upon increasing $H$. For  strong enough anisotropy $A$, the vertical spiral can continuously reduce  the width of the domain wall between spin up and down domains  through the development of higher harmonics. This is the reason why the vertical spiral has lower energy than the conical spiral state. While the multiple-${\bf Q}$ conical spiral is not the global energy minimum for this set of parameters, it always has lower energy than the single-${\bf Q}$ conical spiral. Moreover, in agreement with the variational analysis of Eq.~(\ref{con2}) in Sec.~\ref{sec:Multi-Q Orderings by a Single-Ion Anisotropy}, the multiple-${\bf Q}$ conical spiral becomes
the ground state for a small values of $A \ll 1$.~\cite{leonov2015multiply}

\subsection{Monte Carlo Simulations}
\label{sec:MC}

The MC phase diagrams are obtained by computing  the uniform spin susceptibility, specific heat, and  the spin and chiral structure factors,
\begin{eqnarray}
S^{\nu \nu}_s ( {\bf q}) &=& \frac{1}{N} \sum_{j,l} \langle S^{\nu}_j S^{\nu}_l \rangle e^{i {\bf q} \cdot ({\bf r}_j - {\bf r}_l )},
\nonumber \\
S_\chi^{\mu \mu} ( {\bf q}) &=& \frac{1}{N} \sum_{\gamma, \eta} \langle \chi^{\mu}_{\gamma}  \chi^{\mu}_{\eta} \rangle  e^{i {\bf q} \cdot ({\bf r}_{\gamma} - {\bf r}_{\eta} )},
\label{strf}
\end{eqnarray}
as a function of $H$, $A$ and $T$. The greek labels $\gamma$ and $\eta$ denote the sites of the dual (honeycomb) lattice of the TL.
The brackets $\langle \cdots \rangle$ denote the thermodynamic average. $\chi^{\mu}_{\gamma}={\bf S}_{j} \cdot  {\bf S}_{k} \times  {\bf S}_{l}$ is the scalar chirality on the $\mu =u,d$ (up or down) triangle $jkl$ with center ${\bf r}_{\gamma}$.

\begin{figure}[t!]
\begin{center}
\includegraphics[width=1.0 \hsize]{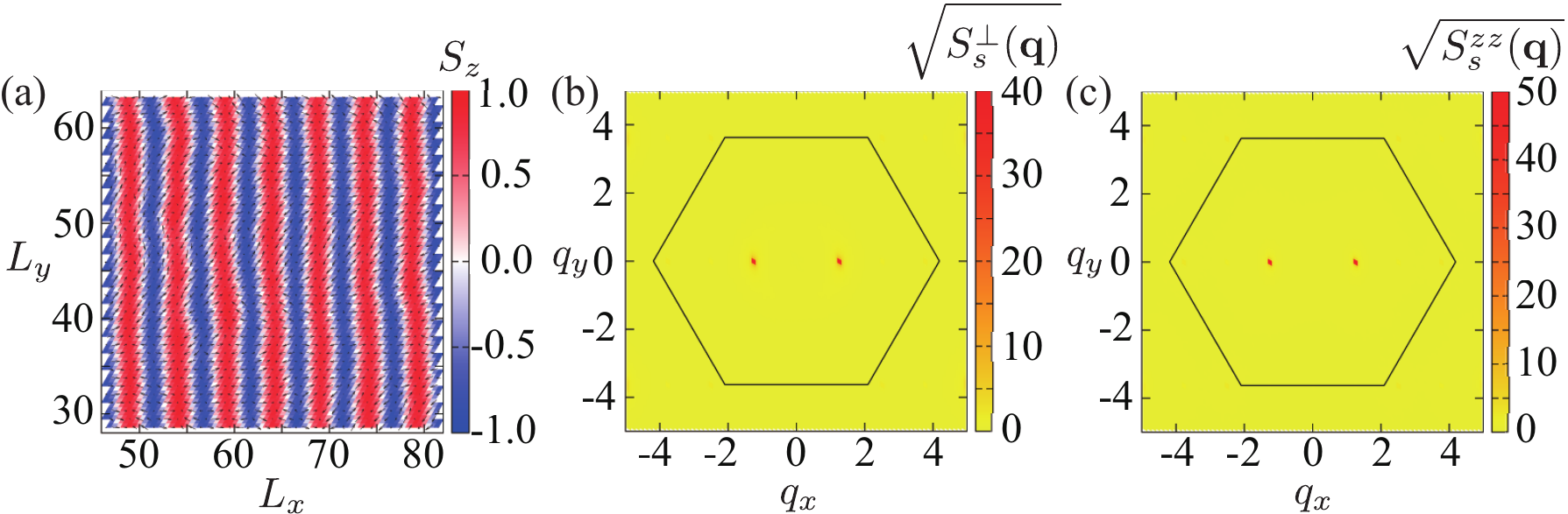} 
\caption{
\label{Fig:data_supple_phases_helical}
(Color online)
Snapshots of spin configurations and the corresponding spin structure factor of the vertical spiral appearing in Fig.~\ref{Fig:phase_smallq}(b) for $Q=2\pi/5$ and $N=100\times 100$. 
The data are taken for $A=0.5$, $H=0.15$, and $T=0.15$. 
In (a), we average over $500$ MCS to integrate out the short wavelength fluctuations.
Panels (b) and (c) show the square root of the $xy$ and $z$ components of the spin structure factor, respectively.  
Here and hereafter, the field-induced $\mathbf{q} = 0$ component is subtracted for clarity.
}
\end{center}
\end{figure}

Figure~\ref{Fig:phase_smallq}(a) shows the $A$-$H$ phase diagram of the $J_1$-$J_3$ model for $L=100$, $J_3/|J_1|=0.5$ and $T=0.18$.  A  conical spiral phase appears for small $A$.  In agreement with our variational analysis of Eq.~\eqref{con2} and direct variational calculations in Fig.~\ref{Fig:fig_supple_variational}, this phase becomes unstable at  lower temperatures.   The vertical spiral phase is induced at low-fields. 
This phase is not a pure single-${\bf Q}$ ordering because of higher harmonics induced by $A$: the optimal vertical spiral is elliptical instead of circular to have the spins more aligned with the easy-axis. 
The real-spin configuration and the spin structure factor of the vertical spiral are shown in Fig.~\ref{Fig:data_supple_phases_helical}. 
As we will discuss in Sec.~\ref{sec:Large-Q}, for larger $A$ values this elliptical distortion eventually evolves into a ``collinear 1D'' phase, which preserves the 1D modulation of the spiral phase.

The skyrmion crystal phase emerges at intermediate magnetic field values and above a rather small critical $A$ value. This phase
narrows down with increasing $A$ because the easy-axis anisotropy naturally favors the fully polarized 
state ($H_{\rm sat}$ decreases with $A$). Except for the second-order phase transition between the conical spiral and the fully polarized state, the other transitions are of  first order, as it is clear from the discontinuities in the magnetization curves, $M(H)$, shown in Fig.~\ref{Fig:phase_smallq}(c). 
We also note that  the magnetization curve has a very small slope ($M \sim 0.4$) inside the skyrmion crystal phase. 

\begin{figure}[t!]
\begin{center}
\includegraphics[width=1.0 \hsize]{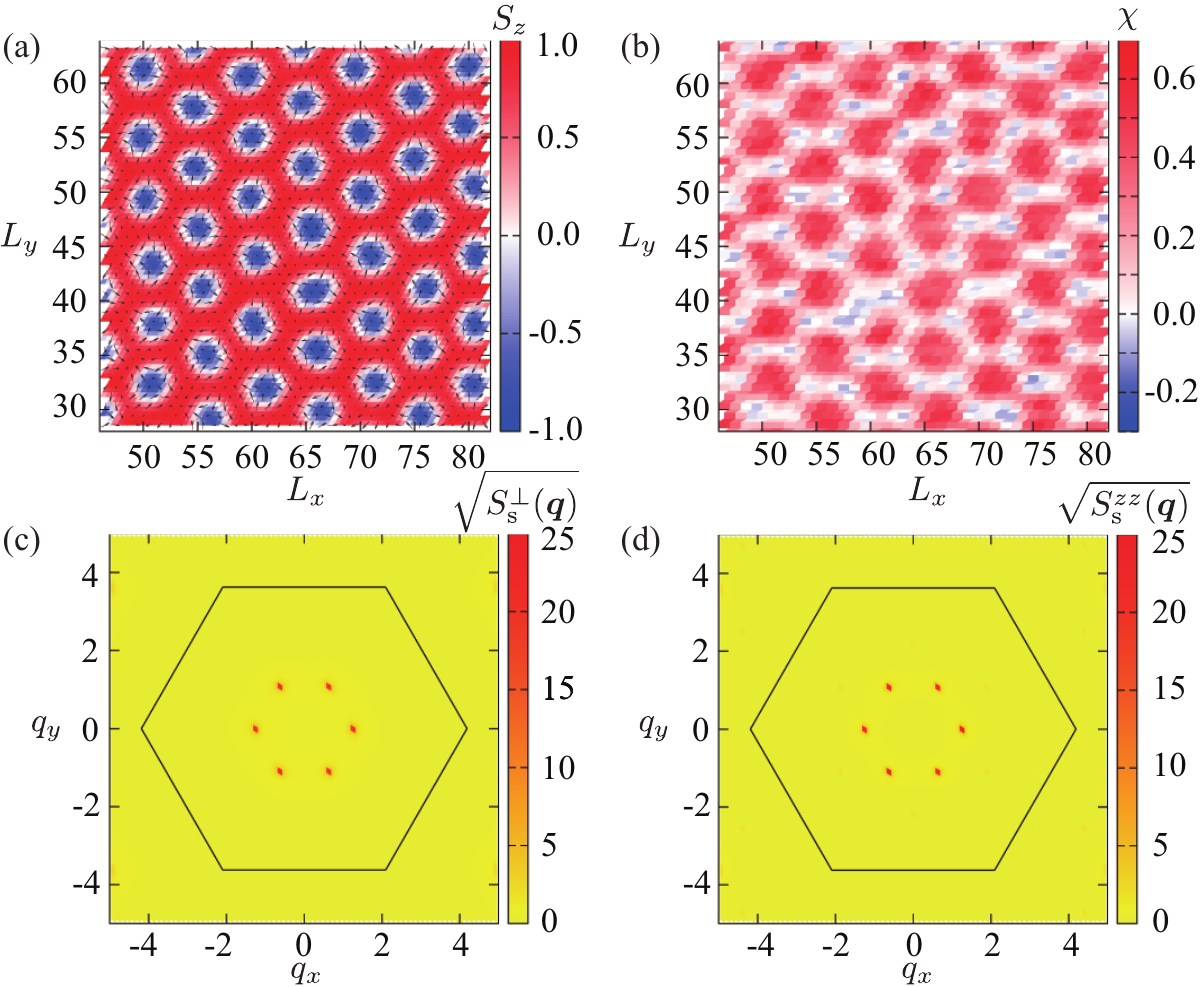} 
\caption{
\label{Fig:dataspin}
(Color online)
Snapshot of the real-space  (a) spin configuration and (b) scalar chirality in the skyrmion crystal phase 
for $H=0.27$ and $T=0.15$ in Fig.~\ref{Fig:phase_smallq}(b). 
The color scale in (a) indicates  the $z$ spin component, parallel to $H$, while the arrows indicate the in-plane $xy$ components. The lattice size is $L=100$ and the MC results are obtained after averaging over $500$ MCS.
Panels (c) and (d) show the $xy$  and $z$ components of the spin structure factor, respectively.  
}
\end{center}
\end{figure}

Figure~\ref{Fig:dataspin}(a) shows a typical real space spin configuration obtained from a snapshot of the MC simulation in the skyrmion crystal phase. The skyrmion cores (blue regions) form a triangular crystal with lattice parameter $4\pi/(\sqrt{3} Q) \sim 5.77$. 
The snapshot of the local scalar chirality, $\chi^{\mu}_{jkl} = {\bf S}_{j} \cdot  {\bf S}_{k} \times  {\bf S}_{l}$, shown in Fig.~\ref{Fig:dataspin}(b), indicates that this phase has a net  uniform scalar chirality,  $\bar{\chi}= \sum_{\langle ijk \rangle}\chi_{ijk}/N$, as expected for a SC. This is confirmed by our finite size scaling analysis of the chiral structure factor in Appendix~\ref{Finite-size Scaling of the Skyrmion Crystal and Other Phases}.
The six peaks in both $S_s^{\perp}$ and $S_s^{zz}$ [see Figs.~\ref{Fig:dataspin}(c) and \ref{Fig:dataspin}(d)] indicate the formation of a triangular SC. Note that $S_s^{\perp}$ can only order at $T = 0$ in 2D,~\cite{Mermin_PhysRevLett.17.1133} while $S_s^{zz}$ can exhibit sharp Bragg peaks at finite-$T$
because the wave-vectors ${\bf Q}_{\nu}$ are commensurate with the underlying TL.

\begin{figure}[b!]
\begin{center}
\includegraphics[width=1.0 \hsize]{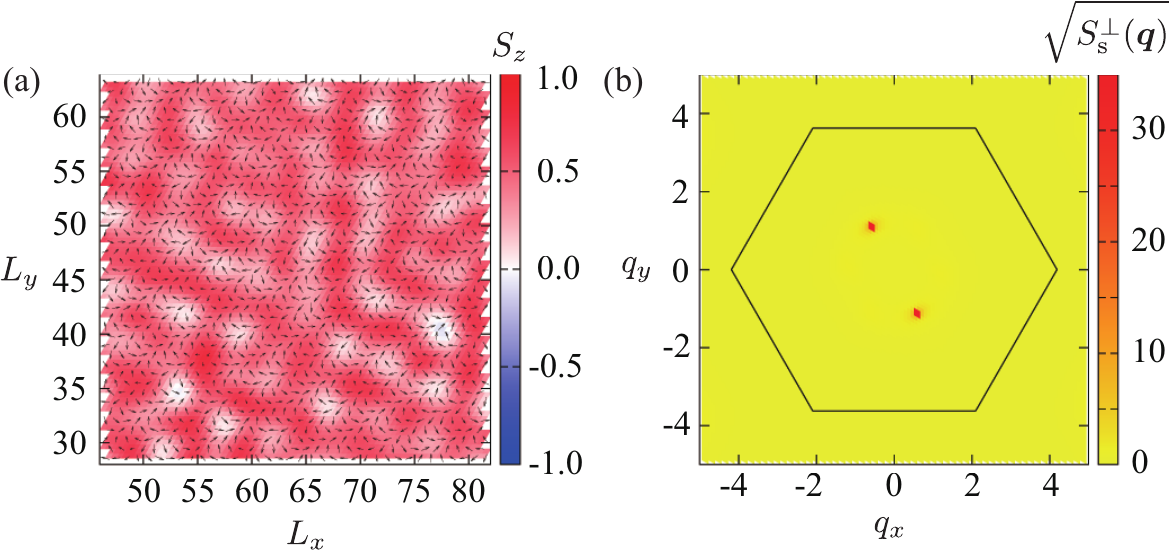} 
\caption{
\label{Fig:data_supple_phases_conical}
(Color online)
(a) Snapshot of the spin configurations and (b) the square root of the $xy$ component of the spin structure factor for the conical spiral phase in Fig.~\ref{Fig:phase_smallq}(a). The Hamiltonian parameters are  $Q=2\pi/5$,  
$A=0.02$, $H=0.43$, $T=0.18$ and the lattice size is $N=75\times 75$.
The snapshot of the spin configuration in (a) is obtained after averaging over $2000$ MCS in order to integrate out the short wavelength fluctuations. The field-induced $\mathbf{q}=0$ component of the spin structure factor has been omitted in (b). 
Note that the $z$-spin component remains uniform in the conical spiral state. 
}
\end{center}
\end{figure}

\begin{figure}[t!]
\begin{center}
\includegraphics[width=1.0 \hsize]{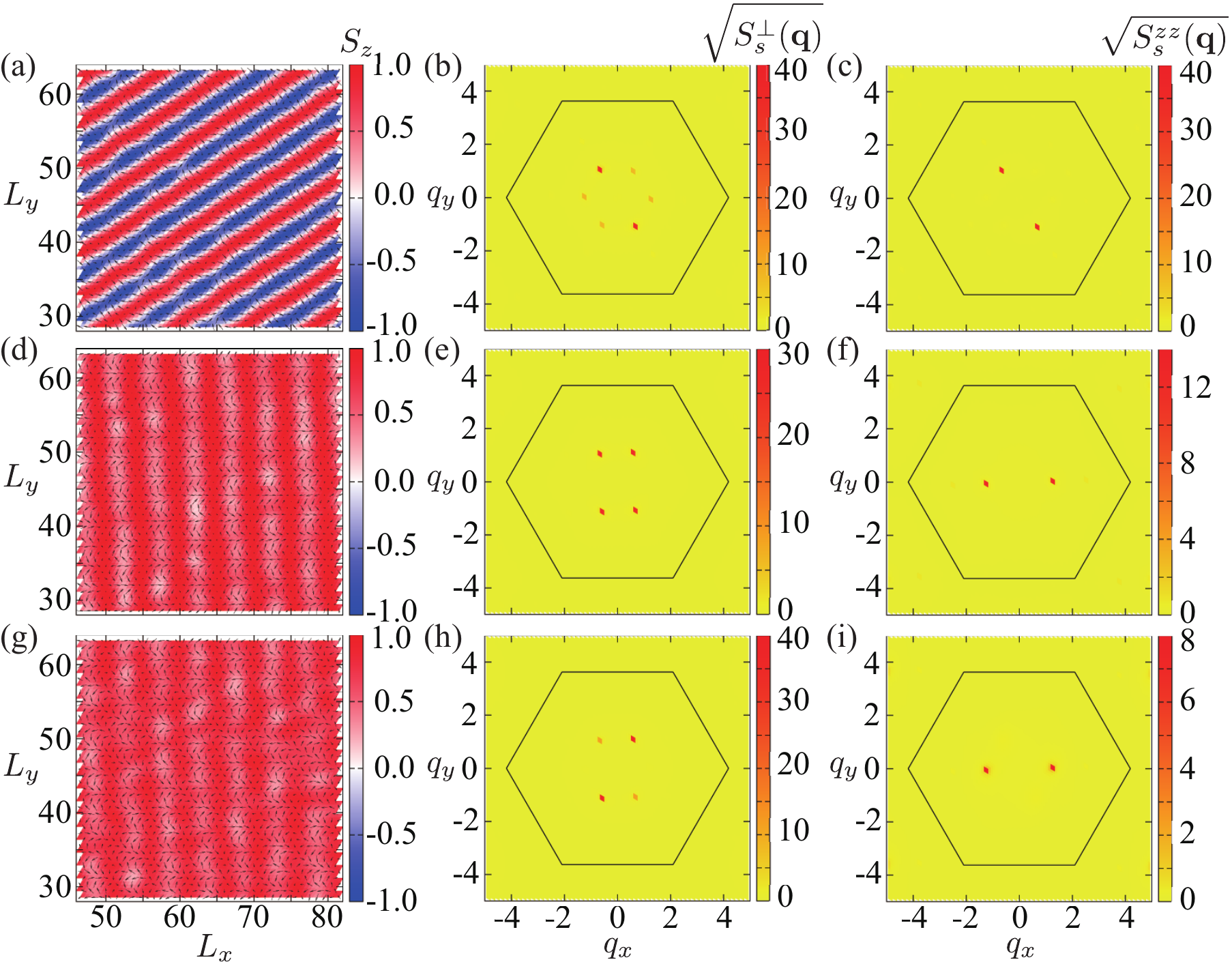} 
\caption{
\label{Fig:data_supple_smallani}
(Color online)
Snapshots of spin configurations and the square root of the spin structure factor for $Q=2\pi/5$, $T=0.02$ and $N=75\times 75$ spins. (a)-(c) Multiple-${\bf Q}$ vertical spiral obtained for $A=0.03$ and $H=0.19$, (d)-(f) multiple-${\bf Q}$ conical spiral with equal transverse amplitudes, $I_x=I_y$, obtained for $A=0.05$ and $H=0.64$,  and (g)-(i) Multiple-${\bf Q}$ conical spiral with different transverse amplitudes, $I_x\neq I_y$,  obtained for $A=0.02$ and $H=0.67$. 
}
\end{center}
\end{figure}

The real space spin configurations of the other two phases, the single-${\bf Q}$ conical and vertical spirals, are shown Figs.~\ref{Fig:data_supple_phases_helical} and \ref{Fig:data_supple_phases_conical}, respectively.
It is interesting to  compare the  finite temperature MC phase diagram shown in Fig.~\ref{Fig:phase_smallq}(a) with 
the $T=0$ variational phase diagram reported in Ref.~\onlinecite{leonov2015multiply}. As shown in Fig.~\ref{Fig:phase_smallq}(a),  the single-${\bf Q}$ conical state and the single-${\bf Q}$ vertical spiral are the only ordered states at $T=0.18$ for small $H$ and $A$. As shown in Fig.~\ref{Fig:data_supple_smallani}
our Monte Carlo results indicate that these single-${\bf Q}$ states evolve into multiple-${\bf Q}$ states upon lowering the temperature. This behavior is consistent with the $T=0$ variational calculations of Ref. ~\onlinecite{leonov2015multiply}. 
However, it is important to emphasize that the  phase diagram becomes qualitatively different in the presence of moderate  thermal fluctuations.
Indeed, at low enough temperatures the single-${\bf Q}_1$ vertical spiral becomes unstable over a finite field interval 
towards a finite spin modulation in the direction perpendicular to the original spin polarization plane. As it is shown in Figs.~\ref{Fig:data_supple_smallani} (a-c), this additional spin modulation has equal intensity for the ${\bf Q}_2$ and ${\bf Q}_3$ components. In addition, the single-${\bf Q}$ conical spiral state becomes a multiple-${\bf Q}$ conical spiral upon lowering the temperature, in agreement with the analysis presented in Sec.~\ref{sec:Multi-Q Orderings by a Single-Ion Anisotropy}  [see Figs.~\ref{Fig:data_supple_smallani} (d-i)].

The phase diagram of Fig.~\ref{Fig:phase_smallq}(a) also exhibits 
a field-induced transition between the vertical spiral and the skyrmion crystal phase for larger values of $A$. This transition can be interpreted in the following way.
When $A$ becomes a significant fraction of $|J_1|$,  the crossover between the spin down and up regions
of a low-energy spin configuration occurs over length scale of order $\sqrt{J_1/A}$. This length can be made much shorter than $2\pi/Q$ in the long wave length limit $Q\ll 1$, i.e.,  we can assume that the boundary between domains with opposite spin alignment is a line with positive tension. The energy of a given state can then be reduced by minimizing the perimeter of the boundary per unit of area. The effect of $H$ on  the vertical spiral is to 
move the up-down boundaries to the right and the down-up boundaries to the left  in order to shrink (expand) the spin down (up) stripes. This implies that the perimeter per unit of area, $P_h/A_h = Q/\pi$, does not depend on the value of the uniform magnetization $M$ along the field direction. In contrast,  the perimeter per unit area of the SC, $P_s/A_s= 3^{1/4} Q \sqrt{1-M} /2 \sqrt{\pi}$, does depend on $M$ because the skyrmion cores shrink as a function of $H$.
We then expect a transition from the vertical spiral to the skyrmion crystal state when $P_h/A_h\approx P_s/A_s$, which leads
to a critical value of $M_{\rm c} \simeq 0.265$.  Given that the transition between both phases is 
of first order, we need to consider the average between the $M$ values right below and above the transition. According to the results shown in  Fig.~\ref{Fig:phase_smallq}(c), this average is  $0.24$, which is  in good agreement with our simple estimate. 

Finally, it is also interesting to study the evolution of the finite-$T$ phase diagram towards $T=0$ when $A$ is 
comparable to $|J_1|$. Figure~\ref{Fig:phase_smallq}(b) shows the $T$-$H$ phase diagram for $L=100$, $J_3/|J_1|=0.5$ and $A=0.5$.
The MC results are complemented with $T=0$ variational calculations in Fig.~\ref{Fig:fig_supple_variational}, whose  phase boundaries, denoted with red triangles in Fig.~\ref{Fig:phase_smallq}(c), deviate slightly from the $T \to 0$ extrapolation of the MC results. 
The  skyrmion crystal  phase extends all the way to $T=0$, in agreement with the variational treatment of Ref.~\onlinecite{leonov2015multiply}.

\section{Large-$\mathbf{Q}$}
\label{sec:Large-Q}

\begin{figure}[b!]
\begin{center}
\includegraphics[width=1.0 \hsize]{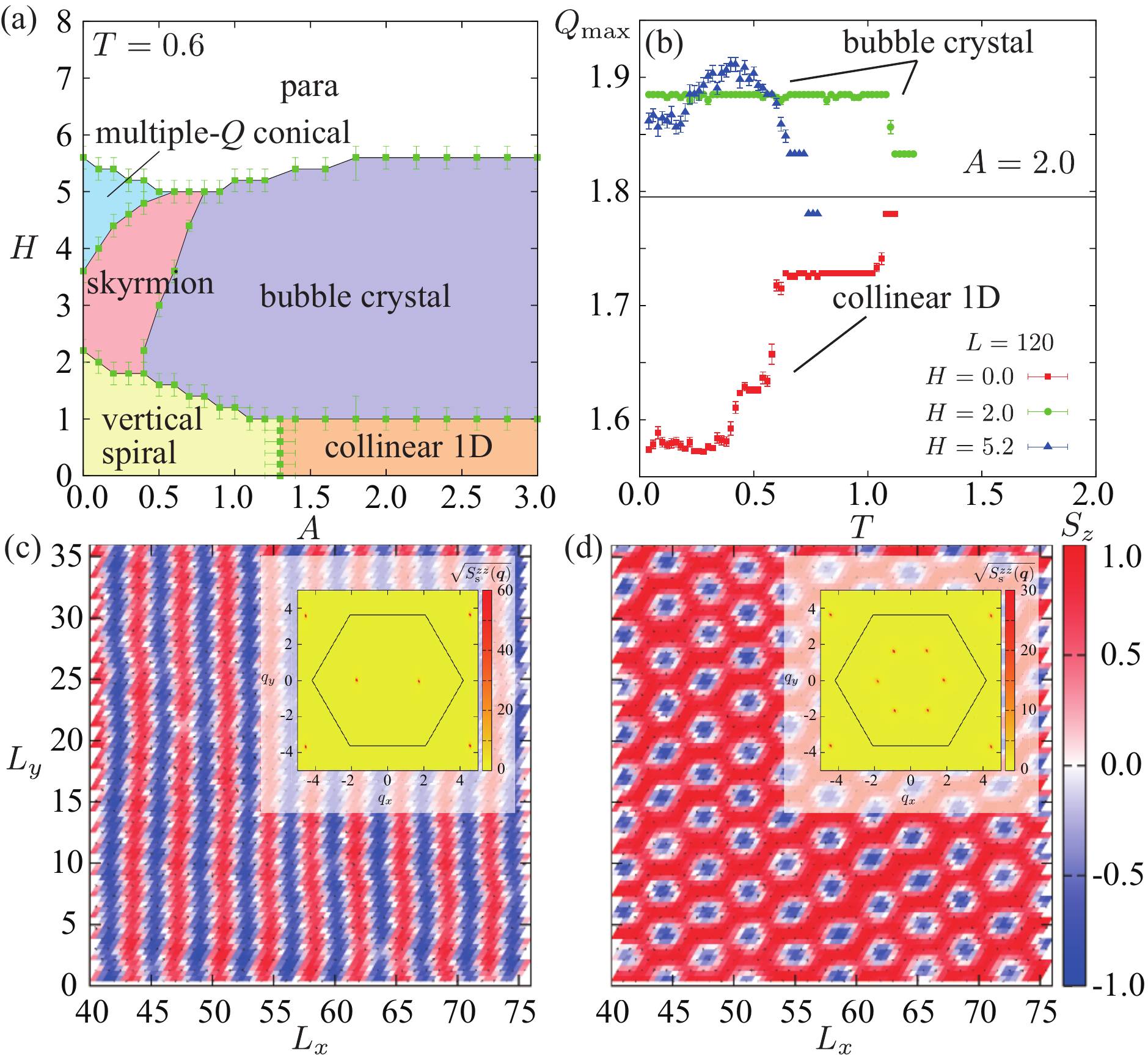} 
\caption{
\label{Fig:phase_largeq}
(Color online)
(a) $A$-$H$ phase diagram of the $J_1$-$J_3$ model at $T=0.60$ and for $Q=4\pi/7$
($J_3 \sim 1.62$). 
(b) $T$-dependence of the ordering-vectors of the collinear 1D and BC phases ($A=2.0$). 
The horizontal line corresponds to $Q=4\pi/7$. 
Panels (c) and (d) show typical spin configurations for the collinear 1D and BC phases, respectively. The insets of panels (c) and (d) show $S^{zz}_s$. 
}
\end{center}
\end{figure}

\begin{figure}[t!]
\begin{center}
\includegraphics[width=1.0 \hsize]{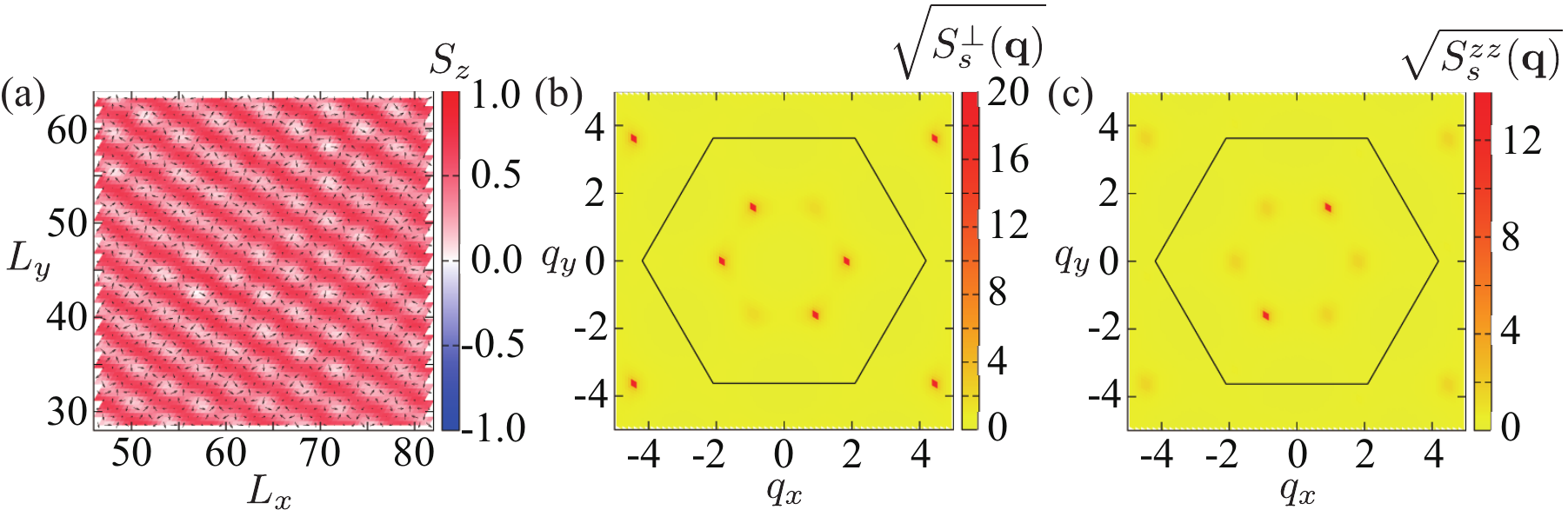} 
\caption{
\label{Fig:data_supple_phases_2q}
(Color online)
(a) Snapshot of spin configuration and (b)-(c) the square root of the spin structure factor of the multiple-${\bf Q}$ conical spiral in Fig.~\ref{Fig:phase_largeq}(a).  The Hamiltonian parameters are $Q=2\pi/5$, $A=0.2$, $H=4.4$, $T=0.6$ and the system size is $N=75\times 75$. 
The snapshot of the spin configuration in (a) is obtained after averaging over $500$ MCS to integrate out the short wavelength fluctuations. The field-induced $\mathbf{q}=0$ component of the structure factor has been omitted in (b) and (c). 
}
\end{center}
\end{figure}

In this Section we will study  the effect of easy-axis spin anisotropy in the large-${\bf Q}$ regime by considering the $J_1$-$J_3$ model with  $Q=4\pi/7$ ($J_3 \sim 1.62$). Figure~\ref{Fig:phase_largeq}(a) shows the typical $A$-$H$ phase diagram at intermediate $T$-values ($T=0.60$) obtained from simulations on lattices of  $N=98 \times 98$ spins. 
Four phases appear in the small $A$ region: vertical spiral, SC, multiple-${\bf Q}$ conical, and paramagnetic states. The vertical spiral and the skyrmion crystal phases are similar to the ones already described for small-$Q$ [see Figs.~\ref{Fig:data_supple_phases_helical} and \ref{Fig:dataspin}]. 
A typical spin configuration for multiple-${\bf Q}$ conical state is shown in Fig.~\ref{Fig:data_supple_phases_2q}. 

Remarkably,  the large $A$ region includes two collinear broken symmetry states. The low-field phase corresponds to a
spin density wave with a 1D modulation, as it is clear from real-space spin configuration shown in Fig.~\ref{Fig:phase_largeq}(c) and from 
the longitudinal spin structure factor, $S^{zz}_s$, shown in the inset of the same figure. In contrast,  the high-field collinear BC phase,
schematically displayed in Fig.\ref{Fig:ponti}(b), is modulated along three principal directions  parallel to the vectors ${\bf Q}_{\nu}$ [see  Fig.~\ref{Fig:phase_largeq}(d) and its inset]. Similar BC phases have been previously discussed in different contexts.~\cite{bobeck1971magnetic,Thiele_PhysRevLett.30.230,Seshadri_PhysRevLett.66.2774}
The local scalar chirality induced by thermal fluctuations
near the phase boundary between the skyrmion and the bubble crystals  decreases gradually and disappears for increasing $A$. Consistently with this behavior, $S^{\perp}_{s}$ exhibits quasi-long range ordering 
in the finite-$T$ skyrmon crystal phase and only short range correlations in the bubble crystal phase.
 
Another interesting aspect of the collinear phases is the temperature dependence of their spatial modulation,
similar to the well-known  case of the axial next-nearest-neighbor Ising (ANNNI) model.~\cite{Elliott_PhysRev.124.346,Fisher_PhysRevLett.44.1502,selke1988annni} 
We note that in both cases there is a competition between frustrated exchange couplings and easy-axis anisotropy. Moreover, the low-field collinear state of Fig.~\ref{Fig:phase_largeq}(a) exhibits a spontaneous 1D modulation similar to the case of the ANNNI model. As expected, the dominant ordering wave-vector
of the low-field collinear phase (obtained from the peak position of $S^{zz}_s({\bf q})$) exhibits plateaus of different sizes and a quasi-continuous behavior in between [see bottom of Figure~\ref{Fig:phase_largeq}(b)],
which is very similar to the result for the ANNNI model.~\cite{selke1988annni}  

The BC phase can be regarded as a multi-${\bf Q}$ extension of the ANNNI physics. 
The bubble density  increases with decreasing temperature, as shown in Fig.~\ref{Fig:phase_largeq}(b) for $H=2.0$ and $5.2$. Once again, the competition between exchange and anisotropy induces temperature driven 
commensurate-incommensurate transitions. In all cases,  the ordering wave-vector 
evolves towards the $Q$-value  selected by the competing exchange interactions  (largest magnetic susceptibility)
upon approaching the transition to the paramagnetic state (see horizontal line in Fig.~\ref{Fig:phase_largeq}(b)).  However, the moments become longitudinally rigid  upon decreasing temperature
forcing the dominant ordering vector to deviate from the optimal $Q$-value at $T=T_c$. 

\section{Summary}
\label{sec:Summary}

In summary, we found that both spatial and easy-axis spin anisotropies stabilize magnetic field-induced skyrmion crystals  in frustrated magnets. Strong six-fold spatial anisotropy induced by a large ordering wave-vector is enough to stabilize a finite temperature skyrmion crystal in isotropic (Heisenberg) frustrated TL magnets.
However, a small easy-axis anisotropy is required to render the skyrmion crystal stable in the long wave length limit. 
The universality of this continuum limit implies that the same is true for any $C_6$ invariant frustrated lattice model,
such as honeycomb or Kagome. Moreover, our variational argument based on Eq.~\eqref{con2}, which holds for arbitrary $Q$, is also valid for any $C_6$ invariant lattice. 
The skyrmion crystal phase is replaced by a collinear crystal of magnetic bubbles for strong enough spatial and easy-axis anisotropies.

Our study underscores the rich multiple-${\bf Q}$ spin textures that emerge from the combination of {\it frustration} and 
{\it anisotropy}. The following three ingredients are enough to obtain field-induced multiple-${\bf Q}$ ordering: (1) $C_6$ symmetry~\cite{szl}, (2) finite $|{\bf Q}|$ ordering due to competing interactions, and (3) easy-axis anisotropy. 
Fe$_x$Ni$_{1-x}$Br$_2$,~\cite{moore1985magnetic} Zn$_{x}$Ni$_{1-x}$Br$_2$,~\cite{day1981neutron} and an Fe monolayer on Ir(111)~\cite{Bergmann_PhysRevLett.96.167203,heinze2011spontaneous} are then candidate materials to exhibit field-induced skyrmion crystal or BC phases. 

\appendix

\begin{acknowledgments}
Computer resources for numerical calculations were supported by the Institutional Computing Program at LANL. 
This work was carried out under the auspices of the National Nuclear Security Administration of the US DOE at LANL under Contract No. DE-AC52-06NA25396 and was supported by the LANL LDRD-DR Program.
\end{acknowledgments}

\section{Finite-size Scaling of the Skyrmion Crystal and Other Phases}
\label{Finite-size Scaling of the Skyrmion Crystal and Other Phases}

\begin{figure}[htb!]
\begin{center}
\includegraphics[width=1.0 \hsize]{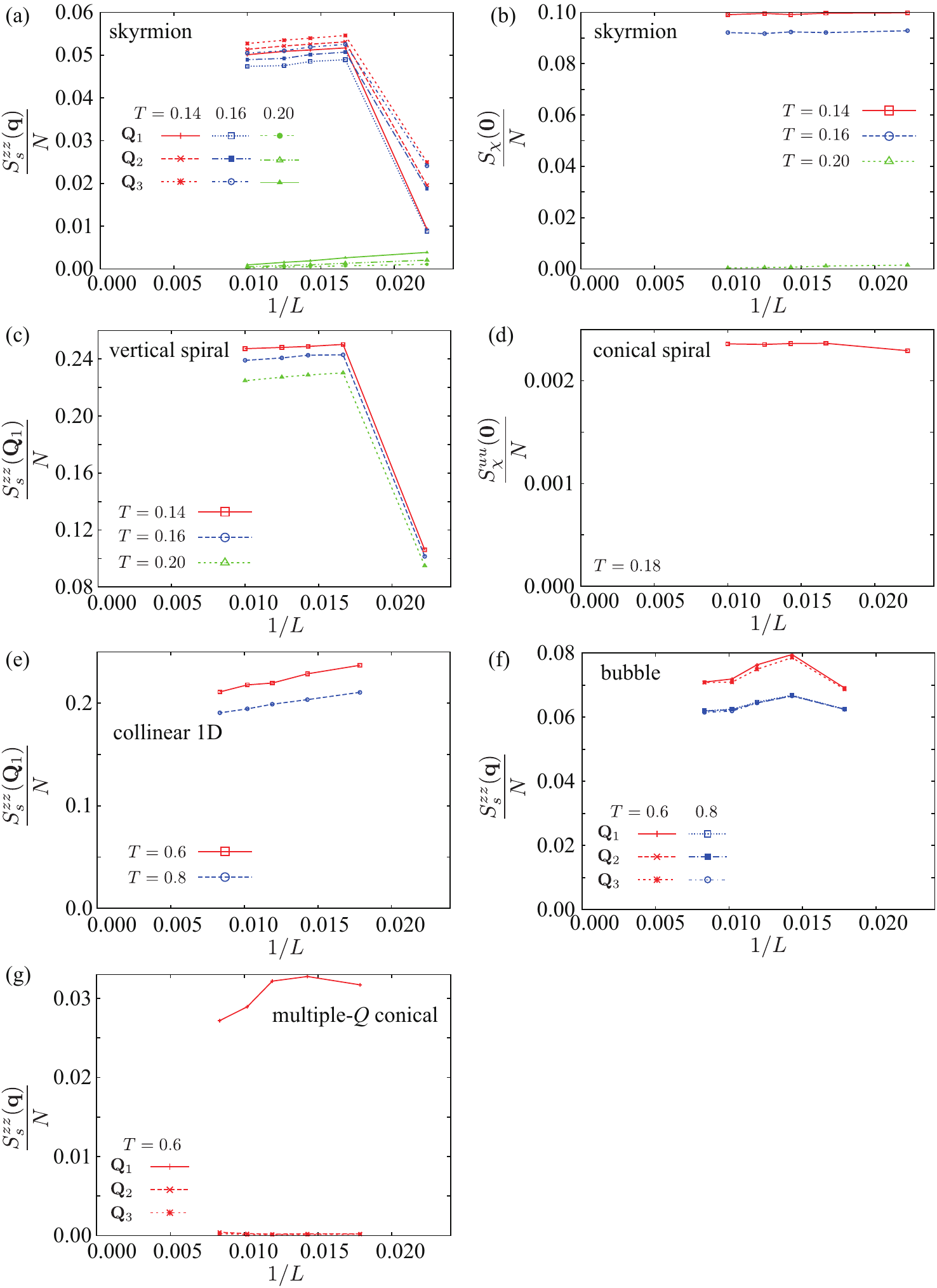} 
\caption{
\label{Fig:sizedep_skyrmion}
(Color online)
Size dependence of the order parameters for each phase: (a) $z$-component of the spin structure factor evaluated at $\mathbf{q}=\mathbf{Q}_1, \mathbf{Q}_2, \mathbf{Q}_3$ and (b) chirality structure factor evaluated at $\mathbf{q}=\mathbf{0}$ ($S_{\chi}(\mathbf{q})=S_{\chi}^{uu}(\mathbf{q})+S_{\chi}^{dd}(\mathbf{q})$) for $Q=2\pi/5$, $H=0.25$, $A=0.5$, and different temperatures in the {\it skyrmion crystal phase}; (c) $z$-component of the spin structure factor evaluated at $\mathbf{q}=\mathbf{Q}_1$  for $Q=2\pi/5$, $H=0.0$, $A=0.5$ in the vertical spiral state; (d) chirality structure factor for the upward triangles of the triangular lattice evaluated at  $\mathbf{q}=\mathbf{0}$ for $Q=2\pi/5$, $H=0.4$ and $A=0.02$ in the {\it conical spiral state}; (e) $z$-component of the spin structure factor with $\mathbf{q}=\mathbf{Q}_1$  for $Q=4\pi/7$, $H=0.0$, and $A=2.0$ in the {\it collinear 1D state}; (f) $z$-component of the  spin structure factor evaluated at $\mathbf{q}=\mathbf{Q}_1, \mathbf{Q}_2, \mathbf{Q}_3$ for $Q=4\pi/7$, $H=2.0$ and  $A=2.0$ in the {\it bubble crystal phase};  (g) $z$-component of the spin structure factor evaluated at $\mathbf{q}=\mathbf{Q}_1, \mathbf{Q}_2, \mathbf{Q}_3$ for $Q=4\pi/7$, $H=4.4$ and $A=0.2$ in the {\it multiple-${\bf Q}$ conical state}. 
}
\end{center}
\end{figure}

In this Appendix we include a finite-size scaling analysis of each phase of the phase diagram shown in Figs.~\ref{Fig:phase_smallq} and \ref{Fig:phase_largeq}. Figures~\ref{Fig:sizedep_skyrmion}(a) and \ref{Fig:sizedep_skyrmion}(b) include the $1/L$ dependence of the 
$z$-component of the spin structure factor and the uniform scalar  chirality normalized by the system size $N$ in the skyrmion crystal phase. 
As expected, the 3-${\bf Q}_{\nu}$ ($\nu=1,2,3$) components of $S^{zz}_s({\bf q})$ extrapolate
to a finite value in the thermodynamic limit ($L \to \infty$). 
The same is true for the uniform scalar spin chirality. 
We also show the finite size scaling analysis  for other phases included in Figs.~\ref{Fig:sizedep_skyrmion}. Panels (c)-(g) include  the vertical spiral in Fig.~\ref{Fig:phase_smallq}(b), the conical spiral in Fig.~\ref{Fig:phase_smallq}(a), the collinear 1D phase  in Fig.~\ref{Fig:phase_largeq}(a), the bubble crystal in Fig.~\ref{Fig:phase_largeq}(a) and the multiple-${\bf Q}$ conical spiral in Fig.~\ref{Fig:phase_largeq}(a) of the main text, respectively.

\bibliographystyle{apsrev}
\bibliography{ref}

\begin{thebibliography}{38}
\bibitem[{szl()}]{szl}
\bibinfo{note}{Skyrmion crystals can also be stabilized in $C_4$ invariant
  lattices with easy-axis anisotropy as long as the $C_4$ spatial anisotropy is
  weak enough. See Ref. [23].}
\end{thebibliography}

\end{document}